\documentclass[prd,twocolumn,aps,dvips]{revtex4}
\usepackage[english]{babel}
\usepackage{amscd}
\usepackage[dvips]{color}
\usepackage{epsfig}
\usepackage{tabularx}
\usepackage{graphicx}
\usepackage{latexsym}
\usepackage{amsmath}
\usepackage{amsfonts}
\usepackage{amssymb}
\usepackage[latin1]{inputenc}

\def\wt{\widetilde}
\newcommand{\sech}{\rm sech}

\usepackage{graphicx}
\usepackage{latexsym}
\usepackage{amsmath}
\usepackage{amsfonts}
\usepackage{amssymb}
\newcommand{\be}{\begin{equation}}
\newcommand{\ee}{\end{equation}}
\newcommand{\ben}{\begin{eqnarray}}
\newcommand{\een}{\end{eqnarray}}
\newcommand{\bes}{\begin{subequations}}
\newcommand{\ees}{\end{subequations}}
\newcommand{\bb}{\bibitem}

\begin{document}
\title{Fermion states on domain wall junctions and the flavor number}
\author{D. Bazeia$^{a}$, F.A. Brito$^{b}$, and R.C. Fonseca$^{a}$}
\affiliation{$^a$Departamento de F\'\i sica, Universidade Federal da
Para\'\i ba, Caixa Postal 5008, 58051-970 Jo\~ao Pessoa, Para\'\i
ba, Brazil
\\
$^b$Departamento de F\'\i sica, Universidade Federal de Campina
Grande, Caixa Postal 10071, 58109-970 Campina Grande, Para\'\i ba,
Brazil}

\date{\today}
\begin{abstract}
In this paper we address the problem of localizing fermion states on stable domain walls junctions. The study focus on the consequences of intersecting six independent 8d domain walls to form 4d junctions in a ten-dimensional spacetime. This is related to the mechanism of relaxing to three space dimensions through the formation of domain wall junctions. The model is based on six bulk real scalar fields, the $\phi^4$ model in its broken phase, the prototype of the Higgs field, and is such that the fermion and scalar modes bound to the domain walls are the zero mode and a single massive bound state, which can be regarded as a two level system, at least at sufficiently low energy. Inside the junction, we use the fact that some states are statistically more favored to address the possibility of constraining the flavor number of the elementary fermions.
\end{abstract}
\maketitle

{\bf I. Introduction.} In recent investigations in high energy physics one of the longstanding 
problem addressed concerns the unveiling of the possible reasons for the Universe to present one time and three spatial dimensions. Specially in \cite{dilute_strings}, the authors suggest that diluting modes in string and D-brane gas cosmology \cite{dilute_branes} favor such Universe. 
An alternative view is given in \cite{Bazeia:2005wt}, in which two of us propose another mechanism of relaxation to three spatial dimensions
through the formation of domain wall junctions. The basic idea is based in the simple fact that d-dimensional objects may intersect one another to form junctions through a dimensional reduction from d to d-1 dimensions, thus giving rise to d-1 dimensional objects. With this idea, and assuming the point of view of Superstring Theory, we suppose that the Universe is early ten-dimensional. In this case, there should be six intersecting 8d domain walls (see below) to form three dimensional junction which evolves in time leading to the worldvolume required by our 4d Universe. The contents of the Universe may then be described through the localization of gravity, gauge and matter fields inside the domain wall junction, an issue which have been considered in many studies in the literature \cite{localiza_junction}.

An interesting issue which has appeared in several investigations concerns the localization of fermion zero modes in order to explain the flavor hierarchy  \cite{flavor_h} of the standard model of particle physics, the so-called $SU(3)\times SU(2)\times U(1)$ model which joins together the strong $SU(3)$ and electroweak $SU(2)\times U(1)$ interactions. In this paper, our main purpose is to extend the former work \cite{Bazeia:2005wt} with the inclusion of fermions, attempting to shed some light in the fact that in our Universe the number of quarks and leptons families seems to be selected to be three. Our results have shown that this number may be related to the number of intersecting domain walls with two bound states, the zero mode and one massive bound state. Interestingly, the presence of the zero mode and a massive bound state is exactly what happens when one investigates a single real scalar field $\phi$ driven by the $\phi^4$ model in its broken phase, the prototype of the fundamental scalar, the Higgs field.

In the present study, we will be mainly interested in the localization of both massive and massless 
fermion states. More specifically, we will consider our system consisting of 8d domain walls
embedded in the 9d+1 space-time, where two fermion bound states are living in due to the partner scalar field being driven by $\phi^4$ model in its broken phase. Six of these domain walls can be joined together to form a stable junction with $2^6$ bound states
-- for sufficiently low energy, there would be no excitation going to the continuum spectrum and so the domain wall can be regarded as a two-level system, the zero mode and a single massive bound state. Thus, if the excited fermion states living on the domain walls have the same energy, say $m$, which we can construct
very naturally, as we will show below, then in the junction made out of six domain walls there would be a distribution of degenerate states that presents a maximum at the energy $\sqrt{3}m$, supposing normal distribution of states. This value suggests that there exist 20 degenerate massive fermion states with this energy that are statistically favored to live in the junction in addition to one non-degenerate fermion zero mode. 

We take advantage of this result to observe that the 21 states might represent six flavors and three colors quark degrees of freedom plus three colorless lepton flavors. One could also think of these 21 fermion degrees of freedom as those to take into account all the matter sector of the Standard Model, given by SU(2) doublets --- the 6 left quarks and 6 left leptons, and SU(2) singlets --- the 6 right quarks and 3 right leptons. According to this scenario anything {\it beyond the Standard Model} should be a manifestation of extra-dimensions. This is because our 21 degrees of freedom are based on the requirement that the higher dimensional physics should effectively manifest as a four-dimensional physics on the junction.

We recall that in the model of \cite{Bazeia:2005wt} one assumes a mechanism of dimensional reduction from the ten-dimensional 
spacetime down to a four-dimensional spacetime which is generated by the domain wall junction. With this, we can then state that
in a 8d domain wall gas in ten dimensions, the chance of a junction to be formed with the superposition of 20 massive states by combining {\it three} 8d domain walls in their fundamental fermion state (the zero mode) and {\it three} 8d domain walls in their excited fermion bound state is statistically favored among any other possible combinations of states.

The study starts in the next Sec.~{II}, where we introduce the model in flat ten-dimensional spacetime and discuss how to construct domain wall solutions and how they can be joined together to form stable junctions. In Sec.~{III} we investigate the presence of fermion states in the junction. Next, in Sec.~{IV} we illustrate the investigations with a simple model, described by two real scalar fields in three dimensions and in Sec.~{V}, we extend the analysis to the more involved case of six scalar fields in $D=9+1$ dimensions. In particular, we study how the scalar fields bind to the junction to form bound states which are partners of the fermion bound states. Finally, in Sec.~{VI} we present our ending comments.\\
\\
{\bf II. The prototype model.} For our purposes in the present study, we restrict ourselves to the fermionic 
and bosonic scalar sectors of a larger supersymmetric theory in ten dimensions \cite{susy_domain}, to find junctions of co-dimension one objects. The number of dimensions is suggested by superstring theory, and so the D8-branes are co-dimension one objects in the $(9,1)$ spacetime dimensions. In this set up, the branes are classical 8d domain wall solutions (8-branes) embedded in 10d spacetime. Let us refer to our theory as a {\it softly broken supersymmetric theory}, in the
sense of Ref.~\cite{Almeida:2001pt}, where a supersymmetric theory is perturbed under the action of a small parameter $\varepsilon$ 
--- here, this parameter is responsible for the stability of the junctions.

Thus the softly broken supersymmetric Lagrangian is written in the following form
\ben\label{Lsusy}
{\cal L}&=&\frac{1}{2}\partial_m\phi^i\partial^m\phi^i+\bar{\psi}^i\Gamma^m\partial_m\psi^i\nonumber
\\
&&+W_{\phi^i\phi^j}\bar{\psi}^i{\psi}^j-V(\phi^i)-\frac{1}{2}\varepsilon F(\phi^i),
\een
where $m=0,1,2,...,D-1$, and $i,j=1,2,...,N$. The scalar potential
is given in terms of the superpotential $W$ by 
\ben\label{scalar_pot}
V(\phi^1,\phi^2...,\phi^N)&=&\frac{1}{2}\frac{\partial
W}{\partial\phi^1}\frac{\partial W}{\partial\phi^1}+\frac{1}{2}\frac{\partial
W}{\partial\phi^2}\frac{\partial W}{\partial\phi^2}+...\nonumber\\
&&+\frac{1}{2}\frac{\partial W}{\partial\phi^N}\frac{\partial W}{\partial\phi^N}.
\een
The scalar fields in the superpotential are such that
\ben
W_{\phi^i\phi^j}=\delta_{ij}W_{\phi^i\phi^i},
\een
and
\ben
V(\phi^1,\phi^2, ...,\phi^N)=V(\phi^1)+V(\phi^2)+...+V(\phi^{N}),
\een 
where $W_{\phi^i\phi^j}$ stands for the second derivative of the superpotential.

The domain wall junctions \cite{junctions} and networks \cite{networks} of domain walls have been adressed in the literature in several contexts. In spite of the difficulty of finding analytical junction solutions, there are known cases in the literarature \cite{exact,networks}. Furthermore, other investigations have been used to address the study of the vacua and energy balance among the instersecting domain walls 
\cite{junctions}, where  individual domain walls are assumed to exist in a way  such that they could be joined together to form a stable junction. These domain walls may form stable junctions, but in this case they should have their tensions satisfying the inequality
\ben
&&|T_{i_1}+T_{i_2}+...+T_{i_N}|<|T_{i_1}|+|T_{i_2}|+...+|T_{i_N}|,\nonumber
\een
where $i_1, i_2,...,i_N=1,2,...,N$ and $T_i=\Delta W_i$ behaves like a vector since it measures $\Delta W$ along straight lines in different directions in the scalar field space $(\phi_1,\phi_2,...,\phi_N)$. In this paper we shall follow the lines of Ref.~\cite{Bazeia:2005wt}, where the individual domain wall tensions were found to satisfy
\ben
&&|T_{i_1}+T_{i_2}+...+T_{i_N}|=|T_{i_1}|+|T_{i_2}|+...+|T_{i_N}|\nonumber
\\&&+\lambda\varepsilon <|T_{i_1}|+|T_{i_2}|+...+|T_{i_N}|,
\een
for $\varepsilon<0$, with $\lambda$ being a positive number which depends on the choice of $F(\phi^i)=\sum_{j>i}^N{F(\phi^i,\phi^j)}$.  To our present purpose it is enough
to use \cite{Bazeia:2005wt}
\ben\label{F}
F(\phi^i,\phi^j)=\frac{1}{2}({\phi^i}^4+{\phi^j}^4)-3{\phi^i}^2{\phi^j}^2+\frac{9}{2}. 
\een

The equations of motion for boson and fermion obtained from the Lagrangian (\ref{Lsusy}) are
\ben\label{eom1}
\square\phi^i+\frac{\partial V}{\partial\phi^i}-W_{\phi^i\phi^i\phi^i}\bar{\psi}^i{\psi}^i+
\frac{\varepsilon}{2}\frac{\partial F}{\partial\phi^i}=0,\\
\label{eom2} \Gamma^m\partial_m\psi^i+ W_{\phi^i\phi^i}\psi^i=0.
\een
By a suitable choice of the superpotential, one finds individual domain wall solutions in the bosonic sector whose dynamics is governed by the equation of motion
\ben\label{phi_eom01}
\square\phi^i+\frac{\partial V}{\partial\phi^i}=0.
\een
For domain wall junctions we are interested in domain wall
solutions that can be joined orthogonally together to form stable junctions. Thus, we shall consider that each scalar field depends on a single spatial coordinate $x^k$, i.e., $\phi(x^1,x^2,...,x^N)\to\phi^k(x^k)\in\{\phi^1(x^1),\phi^2(x^2),...,\phi^N(x^N)\}$,
where $x^k$ is a spatial coordinate transverse to the domain wall described by $\phi^k$. Since each domain wall has co-dimension one, this is very well
defined operation. Under such considerations, static domain walls are governed by the following equations
\ben\label{phi_eom}
\frac{{d}^2\phi^k}{dx_k^2}=\frac{\partial V}{\partial\phi^k}, \qquad k=1,2,...,N.
\een
A first integral of (\ref{phi_eom}) enables us to work with the first-order equations 
\ben\label{phi_eom1st}
\frac{{d}\phi^k}{dx^k}=\frac{\partial W}{\partial\phi^k}, \qquad k=1,2,...,N.
\een
These first-order equations naturally appear in supersymmetric theories and domain walls are BPS solutions preserving 
half of the supersymmetries.

Before going to the next section, some important comments are in order.
To show that a domain wall junction is a mechanism of compactification 
from ten down to four-dimensional spacetime, one should also show that
gravity is localized on it \cite{localiza_junction}. 
This is achieved by considering that the bulk is an $AdS$ spacetime. 
Thus, it is important to show that the bulk
cosmological constant $\Lambda$ in our setup is indeed negative.
Using the fact that the scalar potential in (\ref{scalar_pot}) is
perturbed by the $\varepsilon$-term of (\ref{Lsusy}) we find
\ben V_\varepsilon=\frac12\sum_{i=1}^N(\partial_{\phi_i}W)^2+\frac12\varepsilon
F(\phi^1,\phi^2,...,\phi^N),
\een
where the function $F(\phi^1,\phi^2,...,\phi^N)$
represents $N$ fields combined in the form
$F(\phi^1,\phi^2,...,\phi^N)\!=\!F(\phi^1,\phi^2)+F(\phi^1,\phi^3)+...+F(\phi^{N-1},\phi^N)$. 
Now applying the explicit form of the
superpotential --- see Sec.~IV --- and $F(\phi^i,\phi^j)$ given in Eq.~(\ref{F}), we find that the
perturbed vacua are given by 
$\bar{\phi}_1\!=\!\bar{\phi}_2=\!...\!=\!\bar{\phi}_N\!=\!\pm
\!(3/[2-3(N-1)\varepsilon])^{1/2}$. Thus the bulk cosmological
constant defined as $\Lambda\!\equiv\!V_\varepsilon(\bar{\phi}_1,\bar{\phi}_2,...,\bar{\phi}_N)$ reads
\ben\Lambda=-(27/8)N(N-1)^2\varepsilon^2/[2-3(N-1)\varepsilon],\een for $N>1,$
which is always $negative$ because as we have earlier stressed
$\varepsilon$ should be negative to stabilize the junctions. This result is crucial for the present
investigation, because it circumvents the former result of Ref.~{\cite{nonnorma}}, which stresses
that domain wall junctions in a bulk flat space (where $\Lambda$ should vanish) are not generically able to
localize massless fields on them.\\

{\bf III. The fermion modes.} Let us now investigate the fermion states in the presence of the domain walls
background \cite{zeromodes,nonnorma}. The fermion equation of motion (\ref{eom2}) can be written in terms
of positive and negative energy components $\psi_{\pm}^i$. We look for fermion solutions in the form 
\ben\label{fermions}
\psi^k=e^{ip^{(k)}_\mu x^\mu}\chi^k(x^k),
\een
where $\mu=0,1,2,...,d-2$, are indices labeling coordinates along the domain walls, thus excluding $x^k$ itself from the sum $p^{(k)}_\mu x^\mu$. Now substituting (\ref{fermions}) into (\ref{eom2}), we find
\ben\label{fermions2}
i\Gamma^\mu p^{(k)}_\mu\chi^k-\Gamma^k\partial_k\chi^k+W_{\phi^k\phi^k}\chi^k=0.
\een
Without loosing generality, we take the reference frame where $p^{(k)}_\mu=(E^k,0,...,0)$. In this frame the equation (\ref{fermions2}) takes the simpler form
\ben\label{fermions3}
iE^k\Gamma^0 \chi^k-\Gamma^k\partial_k\chi^k+ W_{\phi^k\phi^k}\chi^k=0.
\een
By using the properties of gamma matrices we have $\Gamma^k\chi_{\pm}=\pm\chi_{\pm}$ and $i\Gamma^0\chi_{\pm}=\chi_{\mp}$, which lead to system of
equations
\bes\ben\label{eqs_ferm}
&&(\partial_k-W_{\phi^k\phi^k})\chi^k_+=E^k\chi^k_-,\\
&&(\partial_k+W_{\phi^k\phi^k})\chi^k_-=-E^k\chi^k_+.
\een\ees
These equations allow us to write the Schr\"{o}dinger-like equations
\bes\label{schall}\ben
&&[-\partial_k^2+U^k_{\pm}(x^k)]\chi^k_{\pm}=E_k^2\chi^k_{\pm},
\\
&&U^k_{\pm}(x^k)=W_{\phi^k\phi^k}^2(x^k)\pm W'_{\phi^k\phi^k}(x^k).
\een\ees
These equations govern the dynamics of the fermion bound states which are linked to the several independent domain walls. To describe domains walls joined together to form a junction we consider the following Schr\"{o}dinger-like equation
\ben\label{sch_junction}
[-\!\nabla^2\!\!+\!U_{\rm junc}]\Psi_{\!\pm(\rm n_1...n_N)}\!\!=\!E^2_{\!\rm(n_1n_2..n_N) junc}\Psi_{\!\pm(\rm n_1...n_N)},
\een
where
\bes\ben
&&U_{\rm junc}=U^1_{\pm}(x^1)+U^2_{\pm}(x^2)+...+U^N_{\pm}(x^N), \\
&& E^2_{\rm (n_1..n_N)junc}={E^2_{(\rm n_1)}}_1\!+\!{E^2_{(\rm n_2)}}_2\!+...+\!{E^2_{(\rm n_N)}}_N, \\
&&\Psi_{\pm(\rm n_1...n_N)}=\chi^1_{\pm(\rm n_1)}(x^1)\times...\times\chi^N_{\pm(\rm n_N)}(x^N),
\een\ees
where the components $\chi^i_{+(\rm n_i)}(x^i)$ and $\chi^i_{-(\rm n_i)}(x^i)$ are normalizable functions. 
We are considering the numbers $n_i=0,1$, i.e., only two bound states, because they can be localized on the individual domain walls -- recall that the scalar sector forms a collection of $\phi^4$ fields in their broken phase. 
The infinite tower of continuum states are non-normalizable states that cannot be localized neither on individual domain walls 
nor on the junction. They would fill the bulk, but they should be seen at higher energy. For the zero modes, ($E=0$), only one of them is normalizable, i.e., the one associated with the chiral fermion on domain walls and junction, as we have learned long ago from Ref.~\cite{zeromodes,nonnorma}.\\
\\
{\bf IV. The two-field example.} Let us consider the example with $N=2$ bulk scalar fields to form $N=2$ independent domain walls to be joined together to form a junction in $D=3+1$ dimensions. The extension of the results to the case of $N$ arbitrary can be done straightforwardly. Consider the following superpotential
\ben\label{W2d}
W(\phi_1,\phi_2)=\lambda_1\left(\frac{\phi_1^3}{3}-a^2\phi_1\right)+\lambda_2\left(\frac{\phi_2^3}{3}-a^2\phi_2\right)
\een
For this case the first-order equations (\ref{phi_eom1st}) reduce to
\ben\label{phi_eom1st_examp}
\frac{{d}\phi^1}{dx^1}=\frac{\partial W}{\partial\phi^1},\;\; \frac{{d}\phi^2}{dx^2}=\frac{\partial W}{\partial\phi^2}.
\een 
There are solutions satisfying these differential equations such as
\ben\label{1st_sol_f}
\phi^1\!(x^1)\!=\!-a\tanh(\lambda_1 ax^1),\;\; \label{1st_sol_q}
\phi^2\!(x^2)\!=\!-a\tanh(\lambda_2 ax^2).
\een
The potentials with upper signs in (\ref{schall}) are given by
\bes\label{Uex}\ben\label{U1ex}
U^1_{+}(x^1)=4\lambda_1^2a^2-6\lambda_1^2a^2{\rm sech}^2{\lambda_1ax^1}, \\
\label{U2ex} U^2_{+}(x^2)=4\lambda_2^2a^2-6\lambda_2^2a^2{\rm sech}^2{\lambda_2ax^2}.
\een\ees
These are modified P\"oschl-Teller potentials \cite{ref00004} of the general form $U(x^{k})=A-B{\,\rm sech\,}^{2}(x^{k})$ for $k=1,2$, with $A$ and $B$ being real constants. The normalizable bound states have the following energies
\ben\label{resul4}
E_{n}=A-\left[\sqrt{B+\frac{1}{4}}-\left(n+\frac{1}{2}\right)\right]^{2},
\een
where
\ben\label{resul4.1}
n=0,1,...<\sqrt{B+\frac{1}{4}}-\frac{1}{2}.
\een
The discrete spectrum is composed by two bound states, the zero mode and the excited state given by
\bes\ben
&&{E_{(0)}^2}_{1,2}=0,\\
&&\chi^{1,2}_{(0)}=C_0{\rm sech}^2{(\lambda_{1,2}ax^{1,2})},\\
&&{E_{(1)}^2}_{1,2}=3\lambda_{1,2}^2a^2,\\
&&\chi^{1,2}_{(1)}=C_1\tanh{(\lambda_{1,2}ax^{1,2})}{\rm sech}{(\lambda_{1,2}ax^{1,2})}.
\een\ees
They are the spectrum of the fermions bound to the domain wall. The spectrum bound to a domain wall junction can be found by using
(\ref{sch_junction}). There are four combinations using zero mode and the excited state that are described by
\ben\label{sch_junction_ex}
&& E^2_{(00)\rm junc}=0, \\
&&\Psi_{(00)}=C_1{\rm sech}^2{(\lambda_{1}ax^{1})}\times{\rm sech}^2{(\lambda_{2}ax^{2})}, \\
&& E^2_{(01)\rm junc}=3\lambda^2_2a^2, \\
&&\Psi_{(01)}\!=\!C_2{\rm sech}^2{(\lambda_{1}ax^{1})}\!\times\!\tanh{(\lambda_{2}ax^{2})}\nonumber\\
&&\;\;\;\;\;\;\;\;\;\;\;\;\times{\rm sech}{(\lambda_{2}ax^{2})},\\
&& E^2_{(10)\rm junc}=3\lambda_1^2a^2, \\
&&\Psi_{(10)}\!=\!C_3\tanh{(\lambda_{1}ax^{1})}\!\times\!{\rm sech}{(\lambda_{1}ax^{1})}\nonumber\\
&&\;\;\;\;\;\;\;\;\;\;\;\;\times{\rm sech}^2{(\lambda_{2}ax^{2})},\\
&&E^2_{(11)\rm junc}=3(\lambda_1^2+\lambda^2_2)a^2, \\
&&\Psi_{(11)}\!=\!C_4\tanh{(\lambda_{1}ax^{1})}\!\times\!{\rm sech}{(\lambda_{1}ax^{1})}\nonumber\\
&&\;\;\;\;\;\;\;\;\;\;\;\;\times\tanh{(\lambda_{2}ax^{2})}\times{\rm sech}{(\lambda_{2}ax^{2})}.\label{sch_junction_ex2}
\een

In the two-field example, the first-order equations (\ref{phi_eom1st_examp}) give the independent BPS domain walls with the kink profiles in (\ref{1st_sol_f}).

Let us now consider the perturbation theory by writing the kink solutions $\phi_s^{k}$ as
a sum of all vibrational normal modes for $k=1,2$, i.e.
\ben\label{pert}
\phi^{k}_{s}(x^k,y^\mu)=\phi^{k}_{s}(x^k)+\sum_{n}\eta^{k}_{n}(x^k)\xi_n(y^\mu),
\een
Here $\mu=0,1,2,3$ stand for the junction world-volume index. Substituting the perturbation (\ref{pert}) into the equations of motion 
(\ref{phi_eom01}), we obtain two Schr\"{o}dinger-like equations for the fluctuations $\eta^{1}_{\rm n_1}(x^1)$ and
$\eta^{2}_{\rm n_2}(x^2)$ that can be written as
\ben\label{MA1}
-\frac{d^2\eta^k_{\rm n_k}}{dx_k^2}+
V_{\rm kj}\eta^k_{\rm n_j}={E^2_{(n_k)}}_k\eta^k_{\rm n_k}, \qquad k=1,2
\een
where we have used $\square\xi_n^k(y^\mu)={E^2_{(n_k)}}_k\xi^k(y^\mu)$. Here $V_{\rm kj}$ are components of the matrix 
\ben\label{Ma3}
V=\left(\begin{array}{cc}
V_{\phi_{1}\phi_{1}} & V_{\phi_{1}\phi_{2}} \\
V_{\phi_{2}\phi_{1}} & V_{\phi_{2}\phi_{2}} \\
\end{array}\right).
\een
For individual domain walls solutions we simply have
\bes\ben
&&V_{\phi_{1}\phi_{1}}=4\lambda_1^2a^2-6\lambda_1^2a^2{\sech^2}{\lambda_1ax^1},
\\
&&V_{\phi_{2}\phi_{2}}=4\lambda_2^2a^2-6\lambda_2^2a^2{\sech^2}{\lambda_2ax^2},
\\
&&V_{\phi_{1}\phi_{2}}=V_{\phi_{2}\phi_{1}}=0.
\een\ees
These potentials were also found for the fermionic case.
Just as in the case of fermions, for each scalar field component, the discrete spectrum is composed of two bound states, the zero mode
and the excited state given by
\bes\ben
&&{E_{(0)}^2}_{1,2}=0,\\
&&\eta^{1,2}_{(0)}=C_0{\rm sech}^2{(\lambda_{1,2}ax^{1,2})},\\
&&{E_{(1)}^2}_{1,2}=3\lambda_{1,2}^2a^2,\\
&&\eta^{1,2}_{(1)}=C_1\tanh{(\lambda_{1,2}ax^{1,2})}{\rm sech}{(\lambda_{1,2}ax^{1,2})},
\een\ees
They are the spectrum of the scalar modes bound to the domain wall. The spectrum bound to a domain wall junction can be found with the use of (\ref{sch_junction}). There are four combinations using the zero mode and the excited state that are described by the same set (\ref{sch_junction_ex})-(\ref{sch_junction_ex2}).

Before closing this section, we should mention that these results in $D=3+1$ can also be useful in other scenarios, for instance, in the cosmological investigations presented in Refs.~\cite{PinaAvelino:2006ia}, where domain wall networks are supposed to fill the spacetime, acting as a possible candidate to describe the dark energy.\\
\\
{\bf V. The four-dimensional model.} We now focus on the extension of the previous results to the case of six scalars in ten-dimensions. We want to find an effective four-dimensional theory for the fields localized on the junction of six orthogonal eight-dimensional domain walls (8-brane) in ten-dimensions. This is required by the framework under consideration, since we are considering $(9,1)$ spacetime dimensions, and the Universe is described by $(3,1)$ dimensions. In this case, each one of the six extra (spatial) dimensions requires a scalar field, leading to the six scalar fields which we will use below. Moreover, since we are considering the Universe as a 3-brane which evolves in time to make its worldvolume a four dimensional spacetime with $(3,1)$ spacetime dimensions, with the 3-brane being a junction of domain walls, we have to impose that all the six scalar fields are immersed
in full ten dimensional space, and then they will form 8d domain walls. They are all co-dimension one global defect structures or
domain walls in the usual sense. 

The Lagrangian for localized fermions states on the 4d junction is given by integrating out the 10d Lagrangian. We start with 
\ben
{\cal L}^F_{4d}=\int{{\cal L}^F_{10d}dx_1dx_2dx_3dx_4dx_5dx_6},
\een
where the fermion dynamics and Yukawa couplings are governed by the Lagrangian
\ben 
{\cal L}^F_{10d}=\bar{\Psi}\Gamma^M\partial_M\Psi+(W_{\phi_1\phi_1}+...+W_{\phi_6\phi_6})\bar{\Psi}{\Psi}.
\een
The scalar and fermion fields are given by the spectral decomposition
\bes\ben
\Phi-\Phi_s&=&\eta(y^\mu;x_1,..,x_6)\nonumber\\
&=&\sum_{n_1...n_6}\!\!\xi^{\rm junc}_{n_1...n_6}(y^\mu)\psi^{n_1...n_6}, \\
\Psi (y^\mu;x_1,..,x_6)&=&\sum_{n_1...n_6}\!\!\tau^{\rm junc}_{n_1...n_6}(y^\mu)\psi^{n_1...n_6},
\een\ees
where $n_i=0,1$ and $\psi^{n_1...n_6}=\chi^{n_1}(x_1)\times...\times\chi^{n_6}(x_6)$, anf the $\chi(x_i)$ are functions that satisfy the equation (\ref{sch_junction}), valid for both fermions and bosons. Since the system has two bound states, there are $2^{N}$ superpartners, i.e., for $N=6$ there are $2^6=64$ four-dimensional scalars $\xi^{\rm junc}_{n_1...n_6}(y^\mu)$ and $2^6=64$ four-dimensional Dirac fermions $\tau^{\rm junc}_{n_1...n_6}(y^\mu)$ living on the junction. Thus we have the four-dimensional Lagrangian
\ben
{\cal L}^F_{4d}&=&\overline{\tau^{\rm junc}_{0...0}}\Gamma^\mu\partial_\mu\tau^{\rm junc}_{0...0}\nonumber\\
&+&\!\!\sum_{n_1...n_6}\overline{\tau^{\rm junc}_{n_1...n_6}}(\Gamma^\mu\partial_\mu-E^{n_1...n_6}_{\rm junc})\tau^{\rm junc}_{n_1...n_6}+ \nonumber\\
&+&\!\!\!\sum_{l_1...l_6}\sum_{m_1...m_6}\sum_{n_1...n_6}g\,\xi^{\rm junc}_{l_1...l_6}\overline{\tau^{\rm junc}_{m_1...m_6}}{\tau^{\rm junc}_{n_1...n_6}}.
\een
Note that the first term describes massless four-dimensional fermions, whereas the second one leads to massive four-dimensional fermions. The Yukawa couplings are controlled by $g$, a constant that is computed by integrating the Yukawa couplings in the six extra dimensions.

For $N$ intersecting domain walls with two bound states, there are $2^{N}$ bound states on the junction. There exists a number of
degenerate states as $\lambda_k=\lambda,$ for any $k$. For $N=6$ intersecting domain walls there is a single state, the zero mode, with vanishing energy. Also, for $m=\sqrt{3}\lambda a$ the other states are given by: 6 states with energy $m$, 15 states with energy $\sqrt{2}m$, 20 states with energy $\sqrt{3}m$, 15 states with energy $\sqrt{4}m$, 6 states with energy $\sqrt{5}m$, and a single state with energy $\sqrt{6}m$. So we have the following distribution:
\ben
(N_f,m)&\!=\!&\{(1,0),(6,m), (15,\sqrt{2}m), (20,\sqrt{3}m),\nonumber\\
&&(15,\sqrt{4}m),(6,\sqrt{5}m),(1,\sqrt{6}m)\}.
\een

Thus the fermions in the Lagrangian have a mass `hierarchy' which goes as follows
\ben
&&{\cal L}^F_{4d}=\bar{\tau}_0^{(0)}\Gamma^\mu\partial_\mu\tau_0^{(0)}\!\!+\!\!\sum_{s=1}^6\sum_{n=1}^{N_s}
\overline{\tau}^{(s)}_{n}(\Gamma^\mu\partial_\mu\!-\!\sqrt{s}\,m)\tau^{(s)}_{n}
\nonumber\\
&&+\sum_{l,l'}\sum_{m,m'}\sum_{n,n'}g_{l'lm'mn'n}\,\xi^{(l')}_{l}\overline{\tau}^{(m')}_{m}{\tau^{(n')}_{n}},
\een
where $N_1=6, N_2=15, N_3=20, N_4=15, N_5=6, N_6=1$ and $l', m', n'=0,1,...,6$.

The partition function for a gas of junctions of six 8d 
domain walls can be found by considering the energy of all fermion states on $6M$
8d domain wall gas in (9+1) dimensions 
\ben
{\wt E}=\sum_{i=1}^{6M}n_i\epsilon_i, \;\;\; \epsilon_i=0,\epsilon,\;\;\; i=1,2,...,6M,
\een
where $\epsilon=3$ and ${\wt E}$ is normalized in the sense of \eqref{resul4}, i.e., in the form ${E^2}/{\lambda^2a^2}$.
Thus the partition function gets the form 
\ben
Z\!=\!\!\sum_{n_1,...,n_{6\!M}}\!\!\exp{\left[-\beta\sum_{i=1}^{6M}n_i\epsilon_i\right]}\!=\!\left[\sum_{n=0}^1\!\exp{(-\beta n \epsilon)}\right]^{6M}\!\!.
\een
The mean energy per domain wall on the junction is given by 
\ben
{\wt u}_{\rm junc}=-\frac{\partial}{\partial \beta}\frac{\ln{Z}}{M}=6\epsilon\frac{e^{-\beta\epsilon}}{1+e^{-\beta\epsilon}}.
\een 
Thus, ${\wt u}_{\rm junc}\to3\epsilon$ at sufficiently high temperature, for $\epsilon/T\ll1$. In this regime the junction energy per domain wall is precisely the same as the energy of {\it three} excited domain walls intersecting {\it three} domain walls in their fundamental state ($\epsilon=0$). As we have mentioned before there are 20 massive states that contribute to the junction energy in this case.

Averaging on the nonzero fermion masses under the distribution for $N_r$ we find
\ben 
<\!m\!>=\frac{\sum_{s=1}^{6} N_s \sqrt{s}\,m}{\sum_{s=1}^{6} N_s}=1.709m\simeq\sqrt{3}m.
\een
This shows that the class of $N_3=20$ distinct fermions with masses $\sqrt{3}m$ is favored. This means that in a domain wall gas in ten dimensions, the probability of a junction to be formed with the superposition of 20 massive states by combining {\it three} 8d domain walls in their fundamental fermion state, and {\it three} 8d domain walls in their single excited fermion state is greater than any other combination. Thus, the observed fermions in our 4d world are governed by the `averaged' Lagrangian
\ben\label{4d_th}
{\cal L}^F_{4d}&\simeq& \bar{\tau}_0^{(0)}\Gamma^\mu\partial_\mu\tau_0^{(0)}\!\!\!+\!\!\!
\sum_{n=1}^{20}\overline{\tau}^{(3)}_{n}(\Gamma^\mu\partial_\mu-\sqrt{3}m)\tau^{(3)}_{n}+\nonumber\\
&&\sum_{l,l'}\sum_{m,m'}\sum_{n,n'}g_{l'lm'mn'n}\,\xi^{(l')}_{l}\overline{\tau}^{(m')}_{m}{\tau^{(n')}_{n}}.
\een 
Assuming that these fermions states can be collected in a vector column which transforms under the local $SU(3)$ group
\ben\label{MA2}
q_{n}=\left(\begin{array}{ccc}
\tau^{1}_{n} \\
\tau^{2}_{n} \\
\tau^{3}_{n} \\
\end{array}\right), 
\een 
we can give $N_c\!=\!3$ colors to {\it six quarks} ($n=1,2,...,6$), i.e., the quark flavor number is $N_F=6$. This comprises $N_cN_F=18$ fermions degree of freedom. There are still two colorless fermions left, that can be put together with the first term (the zero mode) to give rise to {\it three leptons}. Thus, the simple model (\ref{4d_th}) seems to appear as a good approximation to  describe the six quarks and three lepton generations. All the masses are corrected by the Yukawa terms as the scalar fields develop their v.e.v. through the Higgs mechanism -- some highly supressed coupling among quarks and leptons may be acchieved for scalar fields developing  $<\xi^{(l)}_k>\sim0$. A well known example of such coupling concerns the proton decay, where quarks and leptons interact, although we should recall that in the present setup we are not considering four-fermion couplings.

The dynamics of the bosonic modes are described by the Lagrangian
\ben
{\cal L}^B_{10d}&=&\frac{1}{2}\partial_\mu{\eta}\partial^\mu\eta-\frac12\eta(-\nabla^2+U_{\rm junc})\eta-\nonumber\\
&&\frac{1}{3!}\sum_{k=1}^6V^{'''}(\phi_k)\eta^3-\frac{1}{4!}\sum_{k=1}^6V^{''''}(\phi_k)\eta^4.
\een
Integrating out this Lagrangian on the coordinates $x_1,x_2,...,x_6$ we have the four-dimensional Lagrangian
\ben
{\cal L}^B_{4d}&\!=\!&\frac12\sum_{n_1...n_6}\partial_\mu{\xi^{\rm junc}_{n_1...n_6}}\partial^\mu\xi^{\rm junc}_{n_1...n_6}
\!-\!V(\xi),\nonumber\\ 
&\!\simeq\!&\frac12\partial_\mu{\xi^{\rm (0)}_{0}}\partial^\mu\xi^{\rm (0)}_{0}\!\!+\!\frac12\!\sum_{n=1}^{20}\partial_\mu{\xi^{\rm (3)}_{n}}\partial^\mu\xi^{\rm (3)}_{n}\!-\!V(\xi).
\een
The scalar potential is responsible to give non-trivial vacuum solutions to the 21 scalars. They should
be able to give masses to leptons and quarks in the Lagrangian (\ref{4d_th}).\\
\\
{\bf VI. Ending comments.} In this paper we have studied domain wall solutions associated with
six scalar fields $\phi^k(x_k), k=1,2,...,6$. They are responsible for forming a four-dimensional junction with $2^6$ possibly localized modes. However, from the statistical point of view, only 20 degenerate massive states and the zero mode are favored to live on the junction. These 21 localized fermion degrees of freedom are considered to take into account the $N_c=3$ colors and the $N_F=6$ quark flavors quantum numbers, plus three colorless leptons. As we have mentioned before, one could also think of these 21 fermion degrees of freedom as those to take into account all the matter sector of the Standard Model, given by SU(2) doublets --- the 6 left quarks and 6 left leptons, and SU(2) singlets --- the 6 right quarks and 3 right leptons. According to this scenario anything {\it beyond the Standard Model} should be a manifestation of extra-dimensions. This is because our 21 degrees of freedom are based on the requirement that the higher dimensional physics should effectively manifest as a four-dimensional physics on the junction.

In the charicatured setup we suggest that our Universe has selected such quantum numbers because this is statistically favored, for
the Universe being a 3-brane formed as a 3-dimensional junction which evolves in time, with its worldvolume being the standard $(3,1)$ spacetime dimensions. In our opinion, it seems interesting to see how a simple model, suggested in \cite{Bazeia:2005wt} to drive the $(9,1)$ spacetime to form our $(3,1)$ Universe, can capture some important features of the elementary particles, as we see them today.

Although the proposed model is very simple, it seems to unveil some properties which are of current interest. Evidently, at sufficiently higher energies other excitations should appear, in particular the tower of continuum states should be taken into account. However, since they are not bound to the junction, they must live in the bulk and then one should consider them to probe the extra dimensions. The model includes several boson states, which appear in consequence of our starting model, which is based on superstring theory, and up to now we know nothing about their existence. To comply with more realistic models, we should extend the model to add other degrees of freedom. A direct possibility is the inclusion of gauge fields, and we hope to report on this issue in the near future.

We would like to thank CAPES, CNPq, PROCAD-CAPES, and PRONEX-CNPq-FAPESQ for partial financial support.



\begin{thebibliography}{999}

\bibitem{dilute_strings}
  R.~H.~Brandenberger and C.~Vafa,
  ``Superstrings in the Early Universe,''
  Nucl.\ Phys.\  B {\bf 316}, 391 (1989).

  S.~Alexander, R.~H.~Brandenberger and D.~Easson,
  ``Brane gases in the early universe,''
  Phys.\ Rev.\  D {\bf 62}, 103509 (2000)
  [arXiv:hep-th/0005212].

\bibitem{dilute_branes}
  R.~Durrer, M.~Kunz and M.~Sakellariadou,
  ``Why do we live in 3+1 dimensions?,''
  Phys.\ Lett.\  B {\bf 614}, 125 (2005)
  [arXiv:hep-th/0501163].

  A.~Karch and L.~Randall,
  ``Relaxing to three dimensions,''
  Phys.\ Rev.\ Lett.\  {\bf 95}, 161601 (2005)
  [arXiv:hep-th/0506053].


\bibitem{Bazeia:2005wt}
  D.~Bazeia, F.~A.~Brito and L.~Losano,
  ``Relaxing to three dimensional brane junction,''
  Europhys.\ Lett.\ {\bf 76}, 374 (2006)
  [arXiv:hep-th/0512331].

\bibitem{localiza_junction}
  N.~Arkani-Hamed, S.~Dimopoulos, G.~R.~Dvali and N.~Kaloper,
  ``Infinitely large new dimensions,''
  Phys.\ Rev.\ Lett.\  {\bf 84}, 586 (2000)
  [arXiv:hep-th/9907209].

  C.~Csaki and Y.~Shirman,
  ``Brane junctions in the Randall-Sundrum scenario,''
  Phys.\ Rev.\  D {\bf 61}, 024008 (2000)
  [arXiv:hep-th/9908186].

  S.~M.~Carroll, S.~Hellerman and M.~Trodden,
  ``BPS domain wall junctions in infinitely large extra dimensions,''
  Phys.\ Rev.\  D {\bf 62}, 044049 (2000)
  [arXiv:hep-th/9911083].

  I.~Oda,
  ``Localization of matters on a string-like defect,''
  Phys.\ Lett.\  B {\bf 496}, 113 (2000)
  [arXiv:hep-th/0006203].

  N.~Sakai and S.~Tomizawa,
  ``Our world as an intersection of walls and a string,''
  Nucl.\ Phys.\  B {\bf 602}, 413 (2001)
  [arXiv:hep-th/0101042].

  A.~Liam Fitzpatrick and L.~Randall,
  ``Localizing gravity on the triple intersection of 7-branes in 10D,''
  JHEP {\bf 0601}, 113 (2006)
  [arXiv:hep-th/0512247].

\bibitem{flavor_h}
  N.~Arkani-Hamed and M.~Schmaltz,
  ``Hierarchies without symmetries from extra dimensions,''
  Phys.\ Rev.\  D {\bf 61}, 033005 (2000)
  [arXiv:hep-ph/9903417].

  G.~R.~Dvali and M.~A.~Shifman,
  ``Families as neighbors in extra dimension,''
  Phys.\ Lett.\  B {\bf 475}, 295 (2000)
  [arXiv:hep-ph/0001072].

  A.~A.~Andrianov, V.~A.~Andrianov, P.~Giacconi and R.~Soldati,
  ``Domain wall generation by fermion self-interaction and light particles,''
  JHEP {\bf 0307}, 063 (2003)
  [arXiv:hep-ph/0305271].
 

\bibitem{susy_domain}

  G.~W.~Gibbons and N.~D.~Lambert,
  ``Domain walls and solitons in odd dimensions,''
  Phys.\ Lett.\  B {\bf 488}, 90 (2000)
  [arXiv:hep-th/0003197].

  K.~Skenderis and P.~K.~Townsend,
  Phys.\ Lett.\  B {\bf 468}, 46 (1999)
  [arXiv:hep-th/9909070].

  F.~Brito, M.~Cvetic and S.~Yoon,
  ``From a thick to a thin supergravity domain wall,''
  Phys.\ Rev.\  D {\bf 64}, 064021 (2001)
  [arXiv:hep-ph/0105010].

  D.~Bazeia, F.~A.~Brito and J.~R.~S.~Nascimento,
  ``Supergravity brane worlds and tachyon potentials,''
  Phys.\ Rev.\  D {\bf 68}, 085007 (2003)
  [arXiv:hep-th/0306284].

  F.~A.~Brito, F.~F.~Cruz and J.~F.~N.~Oliveira,
  ``Accelerating universes driven by bulk particles,''
  Phys.\ Rev.\  D {\bf 71}, 083516 (2005)
  [arXiv:hep-th/0502057].


\bibitem{Almeida:2001pt}
  C.~A.~G.~Almeida, D.~Bazeia and L.~Losano,
  ``Exploring the vicinity of the Bogomolnyi-Prasad-Sommerfield bound,''
  J.\ Phys.\ A  {\bf 34}, 3351 (2001)
  [arXiv:hep-th/0103166].

\bibitem{junctions}

  E.~R.~C.~Abraham and P.~K.~Townsend,
  ``Intersecting extended objects in supersymmetric field theories,''
  Nucl.\ Phys.\  B {\bf 351}, 313 (1991).

  G.~W.~Gibbons and P.~K.~Townsend,
  ``A Bogomolnyi equation for intersecting domain walls,''
  Phys.\ Rev.\ Lett.\  {\bf 83}, 1727 (1999)
  [arXiv:hep-th/9905196].

  S.~M.~Carroll, S.~Hellerman and M.~Trodden,
  ``Domain wall junctions are 1/4-BPS states,''
  Phys.\ Rev.\  D {\bf 61}, 065001 (2000)
  [arXiv:hep-th/9905217].

  P.~M.~Saffin,
  ``Tiling with almost-BPS-invariant domain-wall junctions,''
  Phys.\ Rev.\ Lett.\  {\bf 83}, 4249 (1999)
  [arXiv:hep-th/9907066].

  D.~Bazeia and F.~A.~Brito,
  ``Tiling the plane without supersymmetry,''
  Phys.\ Rev.\ Lett.\  {\bf 84}, 1094 (2000)
  [arXiv:hep-th/9908090].

  M.~A.~Shifman and T.~ter Veldhuis,
   ``Calculating the tension of domain wall junctions and vortices in
  generalized Wess-Zumino models,''
  Phys.\ Rev.\  D {\bf 62}, 065004 (2000)
  [arXiv:hep-th/9912162].

  D.~Binosi and T.~ter Veldhuis,
  ``Domain wall junctions in a generalized Wess-Zumino model,''
  Phys.\ Lett.\  B {\bf 476}, 124 (2000)
  [arXiv:hep-th/9912081].

  S.~Nam and K.~Olsen,
  ``Domain wall junctions in supersymmetric field theories in D = 4,''
  JHEP {\bf 0008}, 001 (2000)
  [arXiv:hep-th/0002176].


\bibitem{networks}

  D.~Bazeia and F.~A.~Brito,
  ``Bags, junctions, and networks of BPS and non-BPS defects,''
  Phys.\ Rev.\  D {\bf 61}, 105019 (2000)
  [arXiv:hep-th/9912015].

  D.~Bazeia and F.~A.~Brito,
  ``Entrapment of a network of domain walls,''
  Phys.\ Rev.\  D {\bf 62}, 101701 (2000)
  [arXiv:hep-th/0005045].

  F.~A.~Brito and D.~Bazeia,
  ``Network of domain walls on soliton stars,''
  Phys.\ Rev.\  D {\bf 64}, 065022 (2001)
  [arXiv:hep-th/0105296].

  P.~Sutcliffe,
  ``Domain wall networks on solitons,''
  Phys.\ Rev.\  D {\bf 68}, 085004 (2003)
  [arXiv:hep-th/0305198].

  M.~Eto, Y.~Isozumi, M.~Nitta, K.~Ohashi and N.~Sakai,
  ``Webs of walls,''
  Phys.\ Rev.\  D {\bf 72}, 085004 (2005)
  [arXiv:hep-th/0506135].

  M.~Eto, T.~Fujimori, T.~Nagashima, M.~Nitta, K.~Ohashi and N.~Sakai,
  ``Effective action of domain wall networks,''
  Phys.\ Rev.\  D {\bf 75}, 045010 (2007)
  [arXiv:hep-th/0612003].

  M.~Eto, T.~Fujimori, T.~Nagashima, M.~Nitta, K.~Ohashi and N.~Sakai,
  ``Dynamics of Domain Wall Networks,''
  Phys.\ Rev.\  D {\bf 76}, 125025 (2007)
  [arXiv:0707.3267 [hep-th]].

  V.~I.~Afonso, D.~Bazeia, M.~A.~Gonzalez Leon, L.~Losano and J.~Mateos Guilarte,
  ``Constructing networks of defects with scalar fields,''
  Phys.\ Lett.\  B {\bf 662}, 75 (2008)
  [arXiv:0710.5663 [hep-th]]; ``Construction of topological defect networks with complex scalar fields,''
  [arXiv:0805.1086 [hep-th]].

\bibitem{exact}

  H.~Oda, K.~Ito, M.~Naganuma and N.~Sakai,
  ``An exact solution of BPS domain wall junction,''
  Phys.\ Lett.\  B {\bf 471}, 140 (1999)
  [arXiv:hep-th/9910095].

  M.~Naganuma, M.~Nitta and N.~Sakai,
  ``BPS walls and junctions in SUSY nonlinear sigma models,''
  Phys.\ Rev.\  D {\bf 65}, 045016 (2002)
  [arXiv:hep-th/0108179].

  J.~P.~Gauntlett, D.~Tong and P.~K.~Townsend,
  ``Supersymmetric intersecting domain walls in massive hyper-Kaehler sigma models,''
  Phys.\ Rev.\  D {\bf 63}, 085001 (2001)
  [arXiv:hep-th/0007124].

  K.~Kakimoto and N.~Sakai,
  ``Domain wall junction in N = 2 supersymmetric QED in four dimensions,''
  Phys.\ Rev.\  D {\bf 68}, 065005 (2003)
  [arXiv:hep-th/0306077].



\bb{nonnorma}K.~Ito, M.~Naganuma, H.~Oda and N.~Sakai,
  ``Nonnormalizable zero modes on BPS junctions,''
  Nucl.\ Phys.\  B {\bf 586}, 231 (2000)
  [arXiv:hep-th/0004188].

\bibitem{zeromodes}

  R.~Jackiw and C.~Rebbi,
  ``Solitons With Fermion Number 1/2,''
  Phys.\ Rev.\  D {\bf 13}, 3398 (1976).

  D.~B.~Kaplan and M.~Schmaltz,
  ``Domain Wall Fermions and the Eta Invariant,''
  Phys.\ Lett.\  B {\bf 368}, 44 (1996)
  [arXiv:hep-th/9510197].

  J.R. Morris and D. Bazeia,
  ``Supersymmetry Breaking and Fermi Balls''
  Phys. Rev. D {\bf54}, 5217 (1996) [arXiv:hep-ph/9607396].
 
  J.~D.~Edelstein, M.~L.~Trobo, F.~A.~Brito and D.~Bazeia,
  ``Kinks inside supersymmetric domain ribbons,''
  Phys.\ Rev.\  D {\bf 57}, 7561 (1998)
  [arXiv:hep-th/9707016].
  
  D.~Stojkovic, ``Fermionic zero modes on domain walls,''
  Phys.\ Rev.\  D {\bf 63}, 025010 (2001)
  [arXiv:hep-ph/0007343].

  Y.~X.~Liu, L.~D.~Zhang, L.~J.~Zhang and Y.~S.~Duan,
  ``Fermions on Thick Branes in Background of Sine-Gordon Kinks,''
  arXiv:0804.4553 [hep-th].

\bb{ref00004} P.M. Morse and H. Feshbach, \textit{Methods of Theoretical Physics}
(McGraw-Hill, New York, 1953), p.1650.


\bibitem{PinaAvelino:2006ia}
P.P. Avelino, C.J.A. Martins, J.~Menezes, R.~Menezes and J.C.R.E.~Oliveira,
``Frustrated expectations: Defect networks and dark energy,''
Phys.\ Rev.\  D {\bf 73}, 123519 (2006) [arXiv:astro-ph/0602540];
``Defect junctions and domain wall dynamics,''
Phys.\ Rev.\  D {\bf 73}, 123520 (2006) [arXiv:hep-ph/0604250];
``Dynamics of domain wall networks with junctions,''
[arXiv:0807.4442 [hep-ph]].

\end{thebibliography}
\end{document}